# Efficient Table-based Function Approximation on FPGAs using Interval Splitting and BRAM Instantiation

CHETANA PRADHAN, MARTIN LETRAS, and JÜRGEN TEICH,
Hardware/Software Co-Design, Department of Computer Science
Friedrich-Alexander-Universität Erlangen-Nürnberg (FAU)

This paper proposes a novel approach for the generation of memory-efficient table-based function approximation circuits for FPGAs. Given a function $f(x)$ to be approximated in a given interval $[x_0, x_0 + a]$ and a maximum approximation error $E_a$, the goal is to determine a function table implementation with a minimized memory footprint, i.e., number of entries that need to be stored. Rather than state-of-the-art work performing an even sampling of the given interval by so-called breakpoints and using linear interpolation between two adjacent breakpoints to determine $f(x)$ at the maximum error bound, first, we propose three interval-splitting algorithms to reduce the required memory footprint drastically based on the observation that in sub-intervals of low gradient, a coarser sampling grid may be assumed to satisfy the maximum interpolation error bound. Experiments on elementary mathematical functions show that a large fraction in memory footprint may be saved. Second, a hardware architecture implementing the sub-interval selection, breakpoint lookup and interpolation at a latency of just 9 clock cycles is introduced. Third, within each generated circuit design, BRAMs are automatically instantiated rather than synthesizing the reduced footprint function table using LUT primitives providing an additional degree of resource efficiency.

CCS Concepts: • **Hardware** → **Power and energy**; • **Computer systems organization** → **Embedded hardware**.

Additional Key Words and Phrases: FPGA, Approximate Computing, Function Approximation, BRAM

## 1 INTRODUCTION

Approximate computing [21] is a new research field that investigates the trade-off between accuracy, latency, energy [11, 21], and cost of computations. Fig. 1 presents a comparison between approximate computing and conventional computing. Here, approximate computing primes high-performance at the expense of low accuracy. For example, many applications like video and image processing tolerate a certain degree of errors made during acquisition, processing and rendering of images. There already exists a plethora of work on approximate circuit design for basic arithmetic operations such as additions [4, 10, 17], multiplications [2, 16], or divisions [5]. However, much less effort has been invested on efficient implementation of function approximation in hardware including trigonometric, exponential, and logarithmic functions. Here, Taylor series expansions or iterative approximation techniques are well known and often applied, but these also come at either very high resource or latency demands. Tabular representations of functions can serve as an alternative solution in case of small quantization and approximation errors. Indeed, they play an important role in function approximation due to providing constant-time evaluations at the cost of high memory demands. E.g., Matlab/Simulink [13] already offers an available optimization framework named Look-up-table (LUT) Optimizer [14] that computes a function table approximation for a given mathematical function to be approximated within a given interval subject to a maximal approximation error. Notably, this framework also allows to semi-automatically generate code (C++ and VHDL) for tabular function approximations. Unfortunately, when synthesizing this VHDL code to a circuit implementation, e.g., for a Field Programmable Gate Array (FPGA) target, the tables are implemented quite inefficiently by using LUTs structures. Consequently, novel Hardware Description Language (HDL) generation and synthesis techniques are needed for resource-efficient function approximation on modern FPGAs. One contribution of this paper is to instantiate internal so-called Block RAM (BRAM) structures [20]. These BRAMs can even be configured individually in terms of number of entries and bit width of each entry. For example, a so-called BRAM18 block can be alternatively configured to store $16,384$ entries of 1 bit, $8,192$ entries of 2 bits, or up to just $1,024$ entries of 32 bits.

In this realm, this paper presents a fully novel table-based approach for function approximation with contributions to drastically reduce the memory footprint compared to a state-of-the-art method and without any sacrifice in approximation error bounds. Concretely, our contributions are:

- Three *interval splitting algorithms* based on the observation that in sub-intervals of low gradient, a coarser sampling grid may be assumed to satisfy a user-given maximum interpolation error bound $E_a$ at any point $x$ within a given interval $[x_0, x_0 + a)$ of interest. Accordingly, each sub-interval owns an individually optimized spacing between breakpoints. The overall memory footprint is minimized by assigning a coarser breakpoint spacing to sub-intervals with small slope. Only sub-intervals with larger slopes require a fine quantization in order to satisfy $E_a$. The proposed algorithms deliver a partition of the given interval into proper sub-intervals such that $E_a$ is never violated over the whole interval range. It is shown that memory footprint reductions of up to 70 % in average are achieved over tables optimized and generated using the state-of-the-art tool LUT Optimizer by Matlab/Simulink [13].
- An *automated design flow* that uses the interval-based tabular function approximation to generate a hardware description in VHDL automatically. The proposed hardware circuit consists of three units. First, an interval selection circuit determines the index of the sub-interval containing the two breakpoints closest to $x$. Second, a table lookup unit that retrieves the two range values ($y$) of the breakpoints enclosing $x$. Finally, a linear interpolation is performed on these two looked-up values to return the approximation of $f(x)$. The whole architecture (depicted in Fig. 7) performs a function evaluation at a latency of 9 clock cycles.
- Finally, instead of synthesizing the reduced footprint tables using LUTs, BRAMs are instantiated, providing an additional degree of resource efficiency.

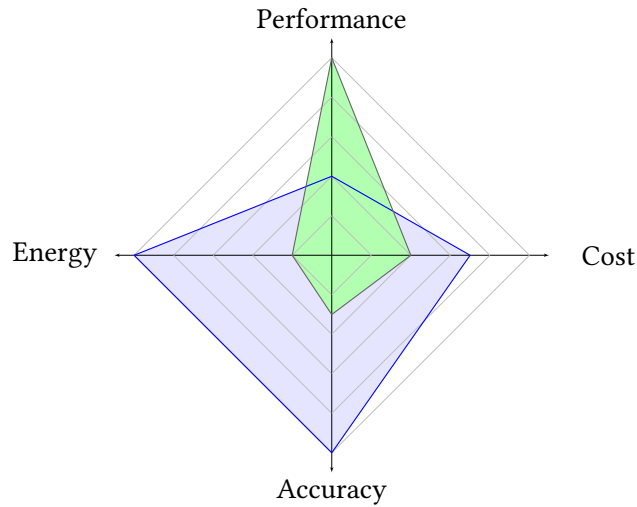

Fig. 1. Approximate computing (green) exploits the error tolerance and robustness of certain applications in order to trade-off accuracy against latency, energy and cost.

The paper is structured as follows: Sec. 2 gives an overview on related work. Then, Sec. 3 presents fundamentals and definitions. Sec. 4 then presents a reference approach according to [13] that will be used for comparison. Subsequently, Sec. 5 introduces the three interval-splitting algorithms and Sec. 6 contains the proposed hardware architecture, and overall design flow. Then, Sec. 7 presents the experimental results. Finally, Sec. 9 concludes our work.

## 2 RELATED WORK

Approximate computing approaches for elementary functions can be classified into three families.
i) Polynomial-based approaches [1, 6, 8, 9, 12, 18] approximate a function by an approximating polynomial. Here, the complexity of the hardware implementation is closely related to the degree of the polynomial. In consequence, polynomial-based approaches tend to be expensive in terms of hardware as high degree polynomials are usually required for achieving low maximal approximation errors.
ii) Iterative approximation algorithms, as the name suggests, calculate a function based on iterative evaluations. Despite being resource-efficient, these algorithms typically suffer from slow convergence, thus requiring many iterations to achieve a certain approximation error. One prominent representative for this type of evaluators are CORDIC algorithms [3, 7], used to approximate trigonometric and hyperbolic functions.
iii) The third family of approximate evaluators is table-based [14, 19]. Table-based approaches split the given interval into a set of discrete points called breakpoints and store the function values evaluated at these breakpoints in a lookup table. Although offering constant-time lookup, the main drawback is often the size of the resulting lookup tables, that grows exponentially with the bit-width of the input. Hence, the table-based method is often combined with the polynomial-based approximation to reduce the memory footprint. This combination usually performs the piecewise-polynomial approximation [6, 8, 9, 12] of the target function.
E.g., similar to our approach, [9] and [6] segment a given interval to be approximated using piece-wise linear interpolations of the form $kx + b$. Both these approaches produce gradient-based non-uniform segments such that the approximation error in each segment does not exceed the specified maximal error. The number of segments determines the number of comparisons performed to place a given input in the correct interval. Both of these numbers increase with the input interval length and the steepness of the function.
Contrary to this, our implementation combines the concept of (1) gradient-based segmentation of the given interval with an input threshold parameter to control the number of segments, and (2) even sampling within a segment to guard a given maximal approximation error bound. Note also that we do not have to store any coefficient at all, but only the range values of the nearest breakpoints $y_i$ and $y_{i+1}$ and we do this in a memory of equal output bit-width for each value.
In this paper, we have shown that the proposed idea of interval-splitting can help to drastically reduce the memory footprint without sacrificing given error bounds.

## 3 FUNDAMENTALS

Storing the range values of a given function $f(x)$ for a discretized domain can help to avoid a computationally expensive polynomial approximation of $f(x)$. Unfortunately, this approach is impractical in case the given approximation error demands fine quantizations of the domain. Indeed, the table size required to store the range values grows exponentially with the bit-width of $x$. E.g., if the bit-width for quantization of $x$ is 8 bits, the size of the lookup table is $2^8 = 255$. However, if the bit-width of $x$ is 32, the lookup table contains $2^{32} = 4,294,967,296$ entries, thus resulting in a memory footprint of 16 GB. Consequently, this method cannot be applied for resource-constrained devices such as FPGAs due to the limited amount of available memory and other hardware resources.

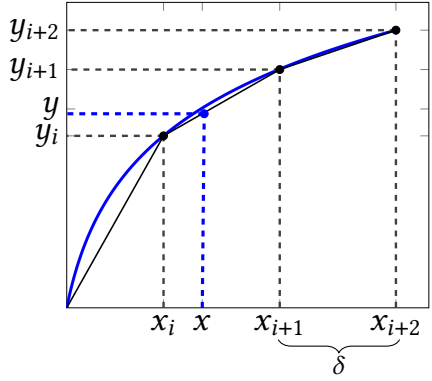

Fig. 2. Piecewise linear interpolation in between two breakpoints $x_i$ and $x_{i+1}$.

As a solution, the combination of tabular function representation and polynomial interpolation has emerged as a widespread technique for function approximation on FPGAs [15]. Instead of storing all the range values of $f(x)$ for a given quantization of $x$, a more efficient approach is to use a coarser quantization of $x$ and perform polynomial approximation in between two adjacent points called breakpoints in the following. In [15], it is proposed to use a degree $n$ polynomial $p(x)$ as follows:

$$p(x) = a_0 + a_1 x + \ldots + a_n x^n \tag{1}$$

Here, $a_i : 0 \leq i \leq n$ denotes the coefficients of the approximating polynomial $p(x)$. Accordingly, the number of entries of the table can be reduced. Still, this requires to store the $n+1$ additional coefficients of each polynomial in between two adjacent breakpoints.

In the rest of this paper, we consider piecewise linear interpolation to avoid the increased computational complexities and memory requirements of higher degree polynomials. In general, breakpoints do not have to be located equidistantly. In this case, linear interpolation can be performed using Eq. (2). For example, the set of $k+1$ breakpoints is denoted as $X = \{x_0, x_1, \ldots, x_k\}$. $Y$ represents the range value of $f(x)$ for $X$ as domain $Y = \{f(x_0), f(x_1), \ldots, f(x_k)\} = \{y_0, y_1, \ldots, y_k\}$. The first and last breakpoints $\{x_0, x_k = x_0 + a\} \in X$ enclose the interval of approximation. Fig. 2 illustrates the calculation of $f(x)$ for any given $x \in \mathbf{R}$. If $x$ matches with any value in $X$, the result can be directly retrieved from a table storing the values of $Y$. Otherwise, the first step is to find the closest breakpoints $x_i$ and $x_{i+1}$ of the input. Their corresponding range values $y_i$ and $y_{i+1}$ are then looked up. Finally, the value of $f(x)$ is computed according to Eq. (2).

$$y = y_i + \frac{x - x_i}{x_{i+1} - x_i}(y_{i+1} - y_i) \tag{2}$$

If $x_{i+1} - x_i$ is not constant $\forall i \in \{0, 1, \ldots, k\}$, then the enclosing breakpoints can only be located by a search method. To this end, the number of comparisons performed grows with the number of breakpoints $k$. This can be avoided by even sampling of the interval $[x_0, x_0 + a)$. Now, when defining a uniform spacing $\delta = (x_{i+1} - x_i) > 0$ between two breakpoints, it is possible to determine the index $i$ describing the interval $[x_i, x_{i+1})$ that includes $x$ as follows:

$$i = \left\lfloor \frac{(x - x_0)}{\delta} \right\rfloor \tag{3}$$

Here, only the first value $x_0$ and the spacing $\delta$ are required to calculate $i$. Finally, with $x_i = x_0 + i \cdot \delta$, Eq. (2) can be re-written to only use the points in $Y$, $x_0$, the location $i$ according to Eq. (3) and the

spacing $\delta$:

$$y = y_i + \frac{x - (x_0 + i \cdot \delta)}{\delta}(y_{i+1} - y_i) \quad (4)$$

Thus, the function approximation can be achieved just by storing the $k + 1$ range values of $X$. Henceforth, we only consider equidistant spacing between two breakpoints. The required memory footprint is referred to as $M_F$ in the following sections. For the above linear interpolation scheme, we obtain a memory footprint $M_F$ of:

$$M_F = k + 1 \quad (5)$$

## 4 REFERENCE APPROACH

In this section, we introduce the *Reference* approach [14] used later (see Sec. 7) for comparison. According to Eq. (5), the memory footprint $M_F$ is linear in $k$, the number of breakpoints. An obvious remaining problem here is how to determine an equi-distant spacing $\delta$ being as large as possible such that still a user-given maximal approximation error bound $E_a$ is never exceeded for any evaluation of $f(x)$ inside the given interval $[x_0, x_0 + a]$. Let $p_i(x)$ denote the linear polynomial used to approximate the function $f(x)$ between the adjacent breakpoints $[x_i, x_{i+1}]$. A two-point line expression can be derived from Eq. (2) as follows:

$$p_i(x) = p_i(x_i) + \frac{x - x_i}{x_{i+1} - x_i}(p_i(x_{i+1}) - p_i(x_i)) \quad (6)$$

If the second derivative $f''(x)$ of $f(x)$ does exist at each point in $[x_i, x_{i+1})$, then the difference between the exact function value $f(x)$ and the value of the approximating polynomial $p_i(x)$ in $x_i \leq x < x_{i+1}$ is given as:

$$f(x) - p_i(x) = \frac{(x - x_i)(x - x_{i+1})}{2} f''(\varepsilon_x) \quad (7)$$

In Eq. (7), $\varepsilon_x$ is some value between $x_i$ and $x_{i+1}$. In consequence, an error bound can be formulated based on Eqs. (6) and (7) as:

$$|f(x) - p_i(x)| \leq \frac{(x - x_i)(x_{i+1} - x)}{2} \max_{x_i \leq x < x_{i+1}} |f''(x)| \quad (8)$$

With the spacing $\delta_i = x_{i+1} - x_i$ between $x_i$ and $x_{i+1}$, the maximum value of $(x - x_i)(x_{i+1} - x)$ in Eq. (8) can be constrained as:

$$\max_{x_i \leq x < x_{i+1}} (x - x_i)(x - x_{i+1}) = \frac{\delta_i^2}{4} \quad (9)$$

By combining Eqs. (8) and (9), we obtain a maximum approximation error bound $E_i$ given a spacing $\delta_i$:

$$E_i = \frac{\delta_i^2}{8} \max_{x_i \leq x < x_{i+1}} |f''(x)| \quad (10)$$

The dependence of the approximation error on the second derivative of a function is intuitive from the perspective of linearity. If $f$ is truly linear in the interval $[x_i, x_{i+1})$, then the second derivative vanishes implying an exact representation. The value of $\max_{x_i \leq x < x_{i+1}} |f''(x)|$ can be expressed in closed form for elementary functions as well as some of the non-elementary functions due to their well-defined second derivatives. Finally, for a given user-defined maximal approximation error bound $E_a$ to hold between any pair of breakpoints and assuming equi-distant spacings $\delta = \delta_i, 0 \leq i \leq k - 1$, we can infer the biggest permissible spacing $\delta$ from the segment $i$ with the smallest value of $\delta_i$ in Eq. (10):

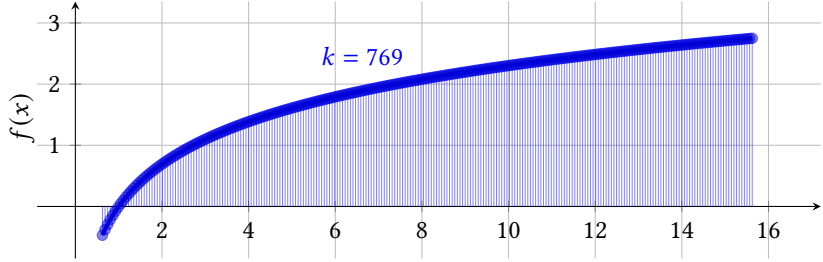

(a) Table-based approximation of $f(x) = log(x)$. Here, $\delta = 0.019$, resulting in a memory footprint of $M_F = k + 1 = 770$ entries.

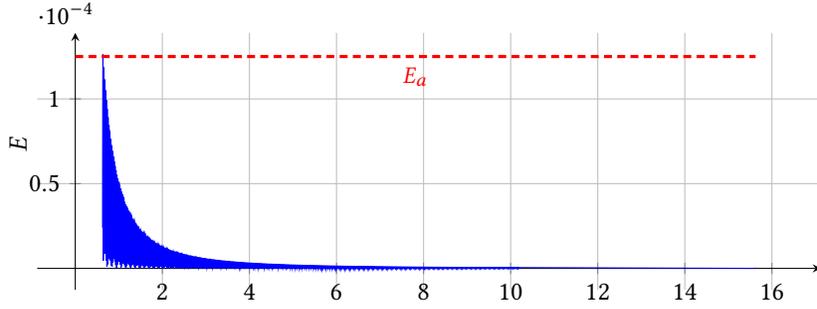

(b) Approximation Error obtained using Eq. (11) given $E_a = 1.25E - 04$.

Fig. 3. Function approximation of $f(x) = log(x)$ in the interval $[0.625, 15.625)$.

$$\delta(f, E_a, [x_0, x_k)) = \min_{0 \leq i \leq k-1} \left( 8 \cdot \frac{E_a}{\max_{x_i \leq x < x_{i+1}} |f''(x)|} \right)^{\frac{1}{2}} \quad (11)$$

E.g., Fig. 3 illustrates the approximation of $f(x) = log(x)$ in the interval $[0.625, 15.625)$ and $E_a = 1.25E - 04$. This results in an spacing $\delta(log(x), E_a, [0.625, 16.625)) = 0.019$. From a spacing $\delta$ and the interval $[x_0, x_0 + a)$, it is possible to generate the entries to be stored in a lookup table (see Fig. 3a). Given a spacing $\delta$ and an interval $[x_0, x_0 + a)$, the memory footprint of this approach called *Reference* approach in the following can be calculated as:

$$M_F^R(\delta, [x_0, x_0 + a)) = \left\lceil \frac{x_0 + a - x_0}{\delta} \right\rceil + 1 \quad (12)$$

Thus, the memory footprint of $f(x) = log(x)$, given the interval $[0.625, 15.625)$, and $\delta = 0.019$ is $M_F^R(0.019, [0.626, 15.625)) = 770$ entries.

The *Reference* approach performs the function approximation given a maximum approximation error $E_a$ by delivering a set of evenly spaced breakpoints. However, this approach does not take into account the gradient in different regions of the interval. This may result in an inefficient footprint, e.g., in Fig. 3, the interval closest to the left-most point $x_0 = 0.625$ determines the maximal spacing for the given approximation error $E_a$. However, it can be seen that for larger values of $x$ (see Fig. 3b), a coarser spacing could be used when splitting the domain into sub-intervals and using different spacings $\delta_i$ within these. Our main idea for the three proposed approaches introduced in the following is therefore to use different spacings $\delta_i$ by splitting the given interval into sub-intervals.

**Algorithm 1:** Binary Interval-Splitter $[p_i, p_{i+1})$

1 **Function** Binary($\omega, E_a, f, p_i, p_{i+1}$)
    **Input:** $f$ is the function to be approximated
           $p_i$ is the lower bound of the interval
           $p_{i+1}$ is the upper bound of the interval
           $\omega$ is the reduction threshold (0,1]
           $E_a$ is the maximum approximation error
    **Output:** $P = \{p_0, p_1, \ldots, p_n\}$ is the set of sub-interval boundaries
2     **begin**
3         $P \leftarrow \{p_i, p_{i+1}\}$       // Initial set of sub-intervals boundaries
4         $\delta_p \leftarrow \delta(f, E_a, [p_i, p_{i+1}))$       // Calculate the spacing in the interval
5         $\kappa_p \leftarrow M_F(\delta_p, [p_i, p_{i+1}))$       // Calculate the $M_F$ of the interval
6         $bp \leftarrow \dfrac{p_i + p_{i+1}}{2}$       // Midpoint between $p_i$ and $p_{i+1}$
7         $\delta_{bp_1} \leftarrow \delta(f, E_a, [p_i, bp))$       // Calculate the spacing of sub-intervals
8         $\delta_{bp_2} \leftarrow \delta(f, E_a, [bp, p_{i+1}))$
9         **if** $\delta_{bp_1} \neq \delta_{bp_2}$ **then**
10             $\kappa_{bp_1} \leftarrow M_F(\delta_{bp_1}, [p_i, bp))$       // Calculate the $M_F$ of sub-intervals
11             $\kappa_{bp_2} \leftarrow M_F(\delta_{bp_2}, [bp, p_{i+1}))$
12             **if** $\kappa_{bp_1} + \kappa_{bp_2} < \kappa_p \cdot \omega$ **then**     // If **true**, accept sub-interval split
13                 $P \leftarrow \{\text{Binary}(\omega, E_a, f, p_i, bp) \cup \text{Binary}(\omega, E_a, f, bp, p_{i+1})\}$
                          // Recursive call for sub-intervals $[p_i, bp)$ and $[bp, p_{i+1})$
14             **end**
15         **end**
16         **return** $P$
17     **end**
18 **End Function**

## 5 INTERVAL SPLITTING ALGORITHMS

Uniform spacing schemes such as even spacing (see the *Reference* approach in Sec. 4) or power of two spacing do not consider the local variation of many functions in the interval of approximation. This section introduces three approaches to partition a given interval $[x_0, x_0 + a)$ into a set of non-uniform sub-intervals and determine a uniform spacing of breakpoints in each sub-interval given a maximum tolerable approximation error $E_a$. The proposed approaches perform the segmentation of the interval $[x_0, x_0 + a)$ to search for interval partitions exploiting the granularity in the spacing according to the steepness of a given function in different sub-intervals. Here, we trade between the number of generated sub-intervals and the memory footprint reduction.

The presented algorithms differ mainly in the heuristic proposed to compute the set of sub-interval partitions. The first two algorithms (Algorithms 1 and 2) split a given interval $[x_0, x_0 + a)$ into two sub-intervals recursively until a corresponding stopping criterion is satisfied. Algorithm 1, called *binary segmentation* always splits a sub-interval at the midpoint. In contrast, Algorithm 2, called *hierarchical segmentation* finds the best splitting point by sweeping over the given interval such that the reduction in memory footprint $M_F$ when splitting is maximized. Here, the step size for the sweep is given as an input. Finally, Algorithm 3 named *sequential segmentation* also performs a sweep-based interval splitting. However here, interval splitting is performed in a single sweep over the overall interval $[x_0, x_0 + a)$.

## 5.1 Binary Segmentation

Algorithm 1 performs a *binary segmentation* of a given interval $[p_i, p_{i+1})$ in which the function $f(x)$ is to be approximated. The inputs of Algorithm 1 are the function $f(x)$, the interval $[x_0, x_0 + a)$, the maximum approximation error $E_a$ and a threshold value $\omega$ which determines whether a new sub-interval split is accepted. *Binary segmentation* determines a partition $P$ of sub-intervals, from which a set of spacings $S$ and a set $K$ containing the numbers of breakpoints in each sub-interval can be obtained. Each element of $p_i \in P$ represents a left (sub-interval) delimiter value. The number of sub-intervals is thus $|P| - 1$. As an example for illustration, assume Algorithm 1 determines the sub-interval splitting $P = \{p_0, p_1, p_2\}$ for a given function, that yields the sets of values $S = \{\delta_0, \delta_1\}$ and $K = \{\kappa_0, \kappa_1\}$. $P$ represents a partitioning of a given interval into two sub-intervals $[p_0, p_1)$ and $[p_1, p_2)$. Here, $\delta_0$ and $\kappa_0$ correspond to the spacing and the number of breakpoints for the sub-interval $[p_0, p_1)$ as well as $\delta_1$ and $\kappa_1$ correspond to sub-interval $[p_1, p_2)$. The memory footprint corresponding to $P$ is therefore given by:

$$M_F^P([p_0, p_{|P|-1})) = \sum_{j=0}^{|P|-1} \kappa_j \qquad (13)$$

*Binary segmentation* is a recursive algorithm, which evaluates the reduction in $M_F$ obtained by splitting the current interval $[p_i, p_{i+1})$ at the midpoint $bp$. Initially, the lower and upper bounds of the interval of approximation ($[x_0, x_0 + a)$) are provided as the inputs $p_i = x_0$ and $p_{i+1} = x_0 + a$ to the function *Binary*. The first step is to initialize $P$ as $\{p_i, p_{i+1}\}$ (see Line 4 in Algorithm 1). The spacing $\delta_p$ and the number of breakpoints $\kappa_p$ are obtained using the *Reference* approach in the interval $[p_i, p_{i+1})$ (see Sec. 4).

The midpoint $bp$ of the input interval is used to create the left sub-interval $bp_1 = [p_i, bp)$ and right sub-interval $bp_2 = [bp, p_{i+1})$. The spacings $\delta_{bp_1}$ and $\delta_{bp_2}$, and the number of breakpoints $\kappa_{bp_1}$ and $\kappa_{bp_2}$ of sub-intervals $bp_1$ and $bp_2$ are then calculated according to Eq. (11) and Eq. (12) (see Lines 8-12 in Algorithm 1). If the sum of $\kappa_{bp_1}$ and $\kappa_{bp_2}$ denoting the memory footprints of $bp_1$ and $bp_2$ is less than a specified fraction (threshold $\omega$) of $\kappa_p$ (see Line 13 in Algorithm 1), the memory footprint reduction and thus split are accepted and the *binary* function is called recursively for the sub-intervals $bp_1$ and $bp_2$, respectively. Otherwise, the algorithm terminates and returns the current input interval boundaries ($\{p_i, p_{i+1}\}$). The final set $P$ returned is the union of all the returned sub-intervals. The value $\omega \in (0, 1]$ indicates the threshold of an acceptable memory footprint reduction. E.g., $\omega = 0.3$ indicates that an interval split must lead to a footprint reduction of at least 30 % of the given interval in order to continue splitting the left and right sub-intervals. Fig. 4 illustrates the proposed recursive sub-interval splitting approximation method using as an example, the function $f = log(x)$ to be approximated in the interval $[p_i, p_{i+1}) = [0.625, 15.625)$ with a maximum approximation error of $E_a = 1.22E - 04$ and splitting threshold $\omega = 0.3$. To find a partition of the interval, *binary segmentation* is performed with the inputs $\omega$, $E_a$, $f$ and $[p_i, p_{i+1})$. Using the *Reference* approach, the uniform spacing $\delta_p$ and memory footprint $M_F$ are obtained as $\delta_p = 0.019$ and $M_F = 770$, respectively. When applying Algorithm 1, the middle point $bp = 8.125$ of the interval $[0.625, 15.625)$ is determined first. The left and right sub-intervals are derived using $bp$, thus $bp_1 = [0.625, 8.125)$ and $bp_2 = [8.125, 15.625)$. After calculating the spacing $\delta_{bp_1} = 0.0195$ and $\delta_{bp_2} = 0.25$ for these two sub-intervals, respectively, the number of breakpoints results in $\kappa_{bp_1} = 384$ and $\kappa_{bp_2} = 31$. The new partition has a memory footprint of $M_F^P = (384 + 31) = 415$. Compared to the memory footprint of 770, the achieved reduction is 46 % which is greater than the required threshold of 30 % ($\omega = 0.3$). Then, the function *binary* is called recursively for the accepted right and left sub-intervals $[0.625, 8.125)$ and $[8.125, 15.625)$, respectively. The previous is repeated until no more splittings can be achieved with an acceptable memory footprint reduction.

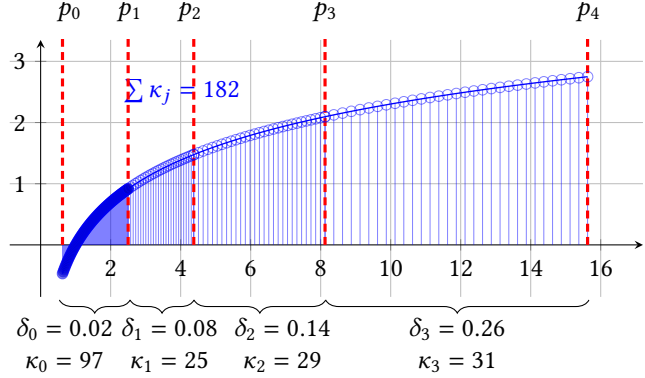

(a) Function approximation of $log(x)$ obtained by *binary segmentation*.

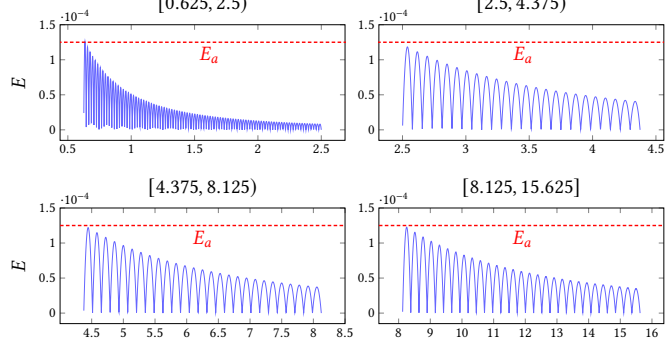

(b) Error margins obtained by *binary segmentation*.

Fig. 4. For a user-given maximal approximation error bound of $E_a = 1.22E - 04$, the algorithm *binary segmentation* identifies the partition $P = \{0.625, 2.5, 4.375, 8.125, 15.625\}$ for splitting threshold $\omega = 0.3$ describing four sub-intervals $[0.625, 2.5)$, $[2.5, 4.375)$, $[4.375, 8.125)$ and $[8.125, 15.625)$. Here, a total of $M_F = 182$ entries represents an overall reduction in memory footprint of 76 % compared to the *Reference* approach (see Fig. 3a).

Algorithm 1 stops with the final partition $P = \{0.625, 2.5, 4.375, 8.125, 15.625\}$. Hence, four sub-intervals $[0.0625, 2.5)$, $[2.5, 4.375)$, $[4.375, 8.125)$, and $[8.125, 15.625)$ were identified (see Fig. 4). The spacings and the number of breakpoints of each sub-intervals are $S = \{0.02, 0.08, 0.13, 0.25\}$ and $K = \{97, 25, 29, 31\}$. From the set of memory footprints $K$, the overall memory footprint $M_F^P([0.625, 15.625)) = (97 + 25 + 29 + 31) = 182$ is calculated. Compared to the reported memory footprint $M_F^R = 770$ for the same function, interval, and error bound (see Fig. 3) using the *Reference* approach, *binary segmentation* is thus able to reduce the memory footprint by 76 % while respecting the maximum approximation error bound imposed by the given $E_a$ (see Fig. 4b).

However, using the midpoint in the partitioning heuristic might lead to a sub-optimal exploration of the partitions in approximation of a given function $f(x)$. The next section introduces an alternative recursive approach which employs a different heuristic in attempt to achieve even larger memory footprint reductions.

## 5.2 Hierarchical Segmentation

*Hierarchical segmentation* is an alternative recursive interval splitting heuristic described by the function *Hierarchical* in Algorithm 2. For each candidate interval $[p_i, p_{i+1})$, a linear sweep is performed instead of just using the midpoint to split the interval. *Hierarchical segmentation* then chooses the splitting point $sp$ with the highest memory footprint reduction and performs a split at $sp$, if the reduction is acceptable. In order to reduce the computational effort, a step size $\varepsilon$ denotes the uniform distance between two adjacent splitting point candidates. The value of $\varepsilon$, the function $f$, the interval $[p_i, p_{i+1})$, the reduction threshold $\omega$, and the maximum approximation error $E_a$ are passed as inputs to Algorithm 2 (see Line 2 of Algorithm 2). The output is the set $P$, from which a set of spacings $S$ and a set of number of breakpoints $K$ are determined (see Line 3 of Algorithm 2). Similar to Algorithm 1, Algorithm 2 initializes the set $P = \{p_i, p_{i+1}\}$ with the lower and upper bounds of a given input interval $[p_i, p_{i+1})$ (see Line 4 in Algorithm 2). The algorithm then computes the maximal number of splitting point candidates $j_{max}$ based on the step size $\varepsilon$ as a parameter (see Line 5 in Algorithm 2). In the next step, we evaluate all the possible memory footprints by splitting the input interval at each candidate point. The splitting point $sp$ is then chosen as the point delivering the smallest memory footprint (see Lines 6 and 7 in Algorithm 2). The left and right sub-intervals are derived as $sp_1 = [p_i, sp)$ and $sp_2 = [sp, p_{i+1})$, respectively. The spacings $\delta_{sp_1}$ and $\delta_{sp_2}$ and number of breakpoints $\kappa_{sp_1}$ and $\kappa_{sp_2}$ of sub-intervals $sp_1$ and $sp_2$ are calculated using the *Reference* approach (see Lines 9-12 in Algorithm 2). If the sum of $\kappa_{sp_1}$ and $\kappa_{sp_2}$ denoting the memory footprint of an interval split at the position $sp$ is less than a specified fraction $\omega$ of $\kappa_p$ (the memory footprint over the interval $[p_i, p_{i+1})$), the split is accepted and the hierarchical function is called recursively to explore the right and left sub-intervals (see Line 13 in Algorithm 2). Otherwise, the algorithm terminates returning the current upper and lower interval bounds (see Line 16 in Algorithm 2). The final set $P$ is then returned as the union of split intervals.

Fig. 5a shows the hierarchical segmentation approach over the interval $[p_i, p_{i+1}) = [0.625, 15.625)$. E.g., let $f(x) = log(x)$, the reduction threshold $\omega = 0.3$, the maximum absolute error $E_a = 1.22E-04$, and the sweep size $\varepsilon = 0.015$. At this point, the number of candidate splitting points is $j_{max} = \frac{15.625 - 0.625}{0.015} = 1000$. This set of points can be expressed in terms of $\varepsilon$ and $j_{max}$ as $\{p_i + j \cdot \varepsilon | i \leq j < j_{max}\}$. After evaluating each candidate, Algorithm 2 determines $sp = 2.90$ as the best candidate resulting in the left sub-interval $sp_1 = [0.625, 2.90)$ and the right sub-interval $sp_2 = [2.90, 15.625)$. $\kappa_{sp_1}$ is equal to 117 and $\kappa_{sp_2}$ is 141. The partition has a memory footprint of $\kappa_{sp_1} + \kappa_{sp_2} = (117 + 141) = 258$ which implies a 66.5 % reduction in memory footprint compared to $M_F^R$ obtained by the *Reference* approach. The achieved reduction is greater than the required threshold reduction of 30 % ($\omega = 0.3$). Thus, the recursive splitting continues for the sub-intervals $sp_1$ and $sp_2$. The previous steps are repeated until no more splitting can be performed resulting in the set $P = \{0.6250, 1.2106, 2.9073, 6.2556, 15.6250\}$, as illustrated in Fig. 5a where the set of spacings $S = \{0.01, 0.06, 0.09, 0.19\}$, and the set of breakpoints $K = \{30, 25, 37, 49\}$. The value of $M_F^P([0.625, 15.625)) = 30 + 45 + 37 + 49 = 161$. Hierarchical segmentation is able to reduce the memory footprint by 79 % with respect to *Reference* approach and by 11.5 % compared to binary segmentation (Algorithm 1).

## 5.3 Sequential Segmentation

Algorithm 3 describes the third approach called *sequential segmentation* and receives as inputs also the function $f(x)$ to approximate, the interval $[x_0, x_0 + a)$, maximum approximation error $E_a$, the reduction threshold $\omega$ and a sweep step size $\varepsilon$. Sequential segmentation performs a linear sweep of the overall input interval similar to the hierarchical segmentation. However, it produces a partition $P$ of a given interval $[x_0, x_0 + a)$ iteratively in a single sweep. The set of splitting point candidates can be represented as $\{sp \mid sp = x_0 + i \cdot \varepsilon\}$ where $1 \leq i < i_{max}$. Here, the distance between two

**Algorithm 2:** Hierarchical Interval-Splitter $[p_i, p_{i+1})$

1 **Function** Hierarchical($\omega$, $E_a$, $f$, $p_i$, $p_{i+1}$)
   **Input:** $f$ is the function to be approximated
   $p_i$ is the lower bound of the interval
   $p_{i+1}$ is the upper bound of the interval
   $\omega$ is the reduction threshold (0,1]
   $E_a$ is the maximum approximation error
   $\varepsilon$ is the sweep step size
   **Output:** $P = \{p_0, p_1, \ldots, p_n\}$ is set of sub-interval boundaries
2  **begin**
3      $P \leftarrow \{p_i, p_{i+1}\}$      // Initial set of sub-intervals boundaries
4      $j_{max} \leftarrow \left\lfloor \dfrac{p_{i+1} - p_i}{\varepsilon} \right\rfloor$      // Calculate the number of splitting point candidates
5      $j^* \leftarrow \underset{1 \leq j \leq j_{max}}{\arg\min} M_F(\delta(f, E_a, [p_i, p_i + j \times \varepsilon)), [p_i, p_i + j \times \varepsilon)) + M_F(\delta(f, E_a, [p_i + j \times \varepsilon, p_{i+1})), [p_i + j \times \varepsilon, p_{i+1}))$
6      $sp \leftarrow p_i + j^* \times \varepsilon$      // Determine the splitting point
7      $\kappa_p \leftarrow M_F(\delta_p, [p_i, p_{i+1}))$      // Calculate the $M_F$ of the interval
8      $\delta_{sp_1} \leftarrow \delta(f, E_a, [p_i, sp))$      // Calculate the spacing of sub-intervals
9      $\delta_{sp_2} \leftarrow \delta(f, E_a, [sp, p_{i+1}))$
10     $\kappa_{sp_1} \leftarrow M_F(\delta_{sp_1}, [p_i, sp))$      // Calculate the $M_F$ of sub-intervals
11     $\kappa_{sp_2} \leftarrow M_F(\delta_{sp_2}, [sp, p_{i+1}))$
12     **if** $\kappa_{sp_1} + \kappa_{sp_2} < \kappa_p \cdot \omega$ **then**      // If **true**, accept interval split
13         $P \leftarrow \{\text{Hierarchical}(\omega, E_a, f, p_i, sp, \varepsilon) \cup \text{Hierarchical}(\omega, E_a, f, sp, p_{i+1}, \varepsilon)\}$
           // Recursive call for sub-intervals $[p_i, sp)$ and $[sp, p_{i+1})$
14     **end**
15     **return** $P$
16 **end**
17 **End Function**

adjacent points in the set is $\varepsilon$ and the number of candidates $i_{max}$ is determined based on the length of interval $a$ and the step size $\varepsilon$ (see Line 2 in Algorithm 3).

Algorithm 3 initializes $P$ and $x_p$ with the lower bound $x_0$ of the input interval (see Lines 1-2 in Algorithm 3). The value of $x_p$ always corresponds to the last entry of the set of partitions $P$. The spacing $\delta_p$ and the memory footprint $\kappa_p$ of the input interval $[x_0, x_0 + a)$ are calculated following the *Reference* approach (Sec. 4). The algorithm iterates through the sweep candidates $sp$ and calculates the memory footprints $\kappa_{sp_1}$ and $\kappa_{sp_2}$ of the sub-intervals $sp_1 = [x_p, sp)$ and $sp_2 = [sp, x_0 + a)$, respectively. If the memory footprint of the input interval $[x_0, x_0 + a)$ split at $sp$ is less than a fraction $\omega$ of $\kappa_p$, then the set $P$ is updated with $sp$ as a new splitting point. The values $x_p$, $\delta_p$ and $\kappa_p$ are updated accordingly (see Lines 13-16 in Algorithm 3). Algorithm 3 stops with the last sweep value which is one step size smaller than $x_0 + a$.

The main difference between *sequential segmentation* and *binary* as well as *hierarchical segmentation* is how the partition points are generated. Here, *sequential segmentation* sweeps from left to right to find the set of partitions, and only one iteration over the given interval is required. In contrast, *binary* and *hierarchical* perform a recursive exploration of each split interval. Every time a new partition point is found, two new sub-intervals located at the right and the left are generated and explored until no more acceptable memory footprint reductions are achieved.

Fig. 5b presents the partition obtained by the algorithm *sequential segmentation* for $f(x) = log(x)$,

**Algorithm 3:** Sequential Interval-Splitter $[x_0, x_0 + a]$

1 **Function** Sequential($\omega, E_a, f, x_0, x_0 + a$)
   **Input:** $f$ is the function to be approximated
       $x_0$ is the lower bound of the interval
       $x_0 + a$ is the upper bound of the interval
       $\omega$ is the reduction threshold (0,1]
       $E_a$ is the maximum approximation error
       $\varepsilon$ is the sweep step size
   **Output:** $P = \{p_0, p_1, \ldots, p_n\}$ is the set of sub-interval boundaries

2   $P \leftarrow \{x_0\}$     // Initialize P with the lower bound of the interval
3   $x_p \leftarrow x_0$
4   $i_{max} \leftarrow \left\lfloor \dfrac{a}{\varepsilon} \right\rfloor$     // Calculate the number of splitting point candidates
5   $\delta_p \leftarrow \delta(f, E_a, [x_p, x_0 + a])$     // Calculate the spacing of sub-interval $[x_p, x_0 + a]$
6   $\kappa_p \leftarrow M_F(\delta_p, [x_p, x_0 + a])$     // Calculate the $M_F$ of sub-interval $[x_p, x_0 + a]$
7   **for** $i \leftarrow 1$ **to** $i_{max}$ **do**
8       $sp \leftarrow x_0 + i \cdot \varepsilon$     // Determine the splitting point
9       $\delta_{sp_1} \leftarrow \delta(f, E_a, [x_p, sp])$     // Calculate the spacing of sub-intervals
10      $\delta_{sp_2} \leftarrow \delta(f, E_a, [sp, x_0 + a])$
11      $\kappa_{sp_1} \leftarrow M_F(\delta_{sp_1}, [x_p, sp])$     // Calculate the $M_F$ of sub-intervals
12      $\kappa_{sp_2} \leftarrow M_F(\delta_{sp_2}, [sp, x_0 + a])$
13      **if** $\kappa_{sp_1} + \kappa_{sp_2} < \kappa_p \cdot \omega$ **then**     // If **true**, accept the interval split
14         $P \leftarrow P \cup \{sp\}$     // Updating P
15         $x_p \leftarrow sp$     // Updating $x_p$
16         $\delta_p \leftarrow \delta(f, E_a, [x_p, x_0 + a])$     // Calc. the spacing
17         $\kappa_p \leftarrow M_F(\delta_p, [x_p, x_0 + a])$     // Calc. the $M_F$ of the interval to split
18      **end**
19   **end**
20   $P \leftarrow P \cup \{x_0 + a\}$     // Updating P with the upper bound of the interval
21 **End Function**

over the interval $[0.625, 15.625)$, the reduction threshold $\omega = 0.3$, maximum absolute error $E_a = 1.22E - 04$, and the sweep size $\varepsilon = 0.3$. Algorithm 3 determines the first splitting point as $sp = 0.9250$ while sweeping the interval $[0.625, 15.625)$. Thus, resulting in the new sub-intervals $sp_1 = [0.625, 0.925)$ and $sp_2 = [0.925, 15.625)$ with a memory footprint of $\kappa_{sp_1} = 16$ and $\kappa_{sp_2} = 510$, respectively. The partition has a memory footprint of $\kappa_{sp_1} + \kappa_{sp_2} = (16 + 510) = 526$ which results in a reduction of 31.6 % compared to the *Reference* approach and is therefore accepted because the memory footprint reduction obtained is greater than the required threshold of 30 % ($\omega = 0.3$). The previous steps are repeated for all the sweep values $sp$ in the interval $[0.625, 15.625)$. Fig. 5b shows that the completed sweep over the interval $[0.625, 15.625)$ produces the partition $P = \{0.625, 0.925, 1.525, 2.425, 3.625, 6.025, 15.625\}$, resulting in a memory footprint $M_F^P([0.625, 15.625)) = (16 + 21 + 19 + 16 + 22 + 52) = 146$. *Sequential segmentation* is able to reduce the memory footprint by 81 % with respect to the *Reference* approach. Compared to the *binary* and *hierarchical* segmentation approaches, the memory footprint reduction is higher by 18 % and 9 %, respectively. Finally, *sequential segmentation* generates six sub-intervals which is larger than the number of sub-intervals produced by the other two heuristics.

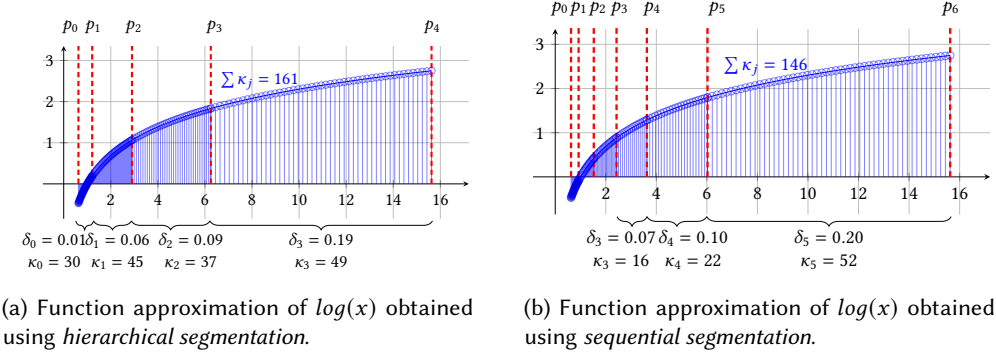

(a) Function approximation of $log(x)$ obtained using *hierarchical segmentation*.

(b) Function approximation of $log(x)$ obtained using *sequential segmentation*.

Fig. 5. Algorithm *hierarchical segmentation* identifies the partition $P = \{0.625, 1.21, 2.90, 6.25, 15.625\}$ for a given splitting threshold of $\omega = 0.3$ describing four sub-intervals $[0.625, 1.21)$, $[1.21, 2.90)$, $[2.90, 6.25)$ and $[6.25, 15.625)$. Here, a total of $M_F = 161$ entries represents an overall reduction in memory footprint of 79 % compared to the *Reference* approach (see Fig. 3a) for a user-given maximal approximation error bound of $E_a = 1.22E - 04$. On the other hand, the *sequential segmentation* obtained the partition $P = \{0.625, 1.21, 2.90, 6.25, 15.625\}$ with $M_F = 146$ resulting in a memory footprint reduction of 81 % compared to the *Reference* approach.

The following section presents an analysis of the proposed approaches to determine if an algorithm delivers significantly better memory footprint reductions than another one.

### 5.4 Comparative Analysis of the Proposed Approaches

As shown by the previous three examples (see Figs. 4 and 5), the proposed partitioning heuristics can achieve significant overall memory footprint reductions by splitting the interval of approximation. This section presents a statistical evaluation of the three presented approaches in terms of the memory footprint reductions delivered compared to the *Reference* approach. We define the memory footprint reduction as:

$$\Delta M_F[\%] = \frac{M_F^R - M_F^P}{M_F^R} \times 100 \qquad (14)$$

In Eq. (14), $M_F^R$ and $M_F^P$ correspond to the memory footprints obtained respectively by the *Reference* approach and any of the three proposed approaches. Our first analysis compares the mean memory footprint reductions of the presented approaches by varying the reduction threshold $\omega$. Six different functions are used as test cases with respective intervals of approximation presented in Table 2.

Fig. 6a shows the memory footprint evaluation of the binary, hierarchical, and sequential segmentation colored in red, gray, and blue. The *x-axis* corresponds to the reduction threshold $\omega$ varying from $\omega=0.01$ to $\omega=0.3$ (1 % to 30 % of memory footprint reductions). The *y-axis* represents the mean memory footprint reduction over a population $X$ of 100 randomly generated intervals contained in the interval $[x_0, x_0+a)$. Table 2 shows the intervals under consideration for each of the test functions. Each point in Fig. 6a represents the mean memory footprint reduction $mean(\Delta M_F)$ obtained by the binary, hierarchical and sequential approaches using the population $X$, a given reduction threshold $\omega$, and maximum absolute error $E_a = 9.5367E - 07$. We can observe the impact of the reduction threshold $\omega$ on $mean(\Delta M_F)$ obtained by the proposed approaches. For small reduction thresholds, all three approaches produce the highest memory footprint reductions and the highest number of generated intervals. Accordingly, a fine-grained exploration of the partitions over the interval is performed. On the other hand, as the value of threshold $\omega$ gets closer to 0.3, the number of generated intervals reduces. Thus, a coarse grain exploration is achieved, and the reduction of memory footprint decreases.

Table 1. Null ($H_0$) and alternate ($H_a$) hypotheses of two-tailed and one-tailed Student's $t$ test

| Tail-type | Null Hypothesis $H_0$ | Alternative Hypothesis $H_a$ |
|---|---|---|
| Two-tailed | $H_0 : \mu_1 = \mu_2$ | $H_a : \mu_1 \neq \mu_2$ |
| Right-tailed | $H_0 : \mu_1 \leq \mu_2$ | $H_a : \mu_1 > \mu_2$ |
| Left-tailed | $H_0 : \mu_1 \geq \mu_2$ | $H_a : \mu_1 < \mu_2$ |

Moreover, for higher $\omega$ values, the sequential segmentation approach (colored in blue) seems to perform better than the other two approaches. For $f(x) = e^{\frac{-x^2}{2}}$ and $f(x) = tan(x)$, the sequential segmentation delivers the highest memory footprint reductions for thresholds $\omega$ close to 0.3 because of the shape of the second derivative (see Eq. (10)), which is irregular and steeper compared with the rest of the test functions.

However, for reduction threshold $\omega$ close to 0 (between 0.01 and 0.04), the hierarchical approach (colored in gray) seems to perform better for almost all the reported test functions. Only for $f(x) = e^x$, the hierarchical and sequential deliver similar reductions. Overall, maximum mean memory footprint reductions using the hierarchical segmentation of 44.5 % for $f(x) = e^x$, 43.5 % for $f(x) = log(x)$, 69.8 % for $f(x) = tanh(x)$, 51.6 % for $f(x) = \frac{1}{1+e^{-x}}$, and 65.8 % for $f(x) = e^{\frac{-x^2}{2}}$ were achieved.

Next, in order to verify the superiority of one algorithm over others, we perform a statistical analysis using a two-sample Student's $t$-test. This statistical test is used to compare the means of two independent groups of samples. The samples are assumed to be normally distributed with an unknown variance. The types of $t$-tests and the corresponding null ($H_0$) and alternative hypotheses ($H_a$) are given in Table 1. A test decision accepts or rejects a hypothesis with a confidence level of $1 - \alpha$ depending on the absolute value of the statistics, the significance level $\alpha$ (usually set to 0.05), the sample size, and the population means ($\mu_1$ and $\mu_2$) of the two compared populations (henceforth groups $G_1$ and $G_2$). In our case, we compare pairwise the mean memory footprint reductions $\Delta M_F$ produced by the three proposed methods. The null hypothesis $H_0$ must be accepted in the right-tailed $t$-test and rejected in the left-tailed $t$-test to establish that the samples in group $G_2$ outperform those in group $G_1$. To perform the one-tailed two-sample $t$-test, we employed the Matlab [13] function $ttest2(G_1, G_2)$ from the Statistics and Machine Learning Toolbox. The output of $ttest2(G_1, G_2)$ is 1 if the null hypothesis $H_0$ is rejected and 0 otherwise. Accordingly, the method represented by $G_2$ yields more significant average memory footprint reductions than the method defined by $G_1$ if the values returned by the right and left tailed $t$-test are 0 and 1, respectively.

For the $t$-test, the considered groups are the mean memory footprint reductions ($mean(\Delta M_F)$) that we have previously produced and visualized in Fig. 6a. Groups $G_1$, $G_2$ and $G_3$ corresponds to binary, hierarchical and sequential approaches, respectively, each with a sample size of 30. Each sample in a group represents the average memory footprint reduction achieved by one of the proposed methods for 100 different input intervals and a given threshold reduction $\omega$. To obtain all the samples in a group, $\omega$ is varied between 0.01 and 0.3, in steps of 0.01. The set of input intervals remains the same for evaluation of the samples in any group. The results of the one-tailed two-sample $t$-tests are presented in Table 2. We can observe that the $t$-test is not statistically conclusive for $f(x) = log(x)$ and $f(x) = e^x$. Here, the confidence level does not allow us to conclude one algorithm significantly outperforming the others in terms of achievable memory footprint reductions. We might say that

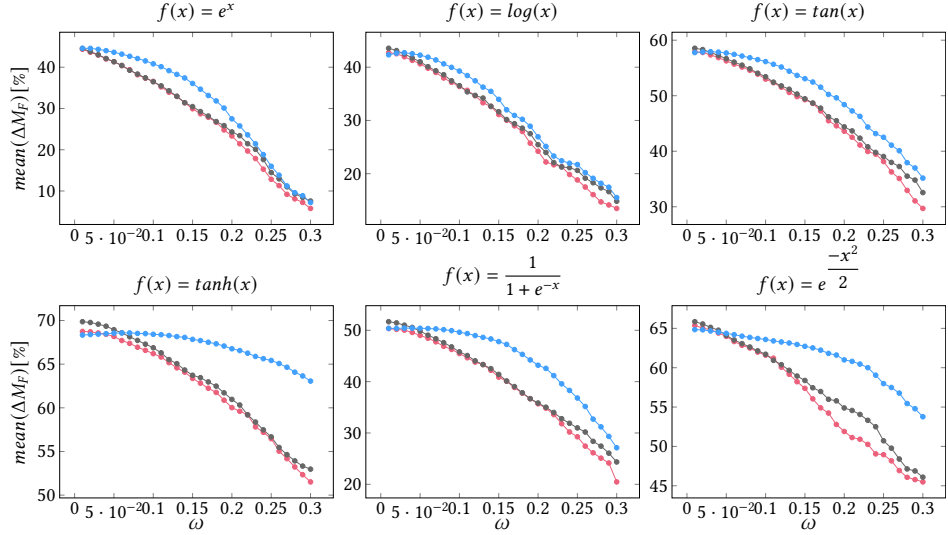

(a) Memory footprint reduction analysis of the proposed binary, hierarchical and sequential approaches.

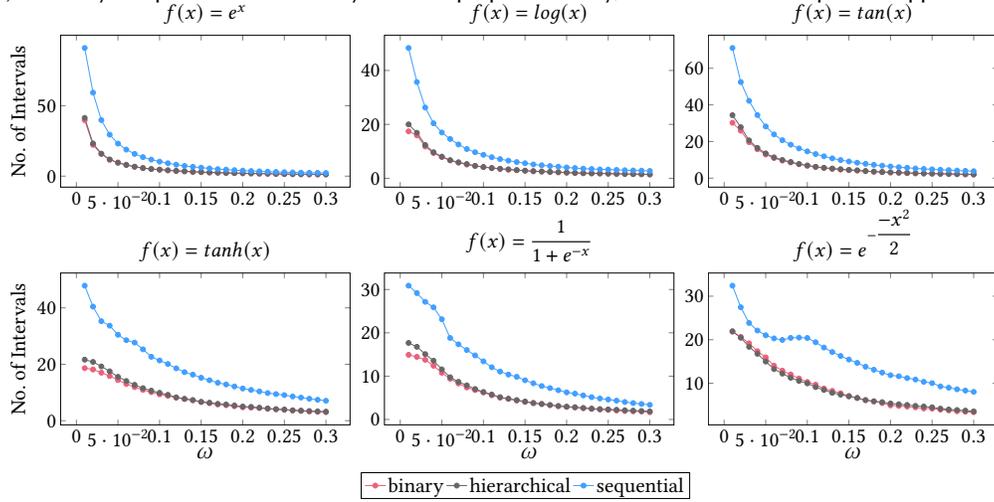

(b) Number of intervals obtained by the proposed binary, hierarchical and sequential approaches.

Fig. 6. Memory footprint reduction analysis of the proposed binary, hierarchical and sequential interval-based segmentation approaches colored in red, gray, and blue, respectively. The *x-axis* presents 30 threshold values $\omega$ ranging from 0.01 to 0.3. In contrast, each point represents the mean memory footprint reduction $mean(\Delta M_F)$ for 100 randomly generated sub-intervals in the interval range presented in Table 2 for each given value $\omega$.

the three approaches deliver similar reductions. However, for $f(x) = tan(x)$, we can conclude that the sequential approach outperforms the binary segmentation, whereas the comparison between sequential and hierarchical $t$-test is non-conclusive. For the other functions in Table 2, we can conclude that the sequential segmentation outperforms the binary and hierarchical approaches with a confidence level of 95 % (see rows colored in gray and the values 0 and 1 for the right-tailed and left-tailed $t$-test, respectively).
Following the results of the $t$-test, we may conclude that the interval splitting heuristic based on sequential segmentation delivers the maximum mean memory footprint reductions. However, the hierarchical segmentation seem to have greater sample mean reductions even compared to the sequential segmentation when the reduction threshold value $\omega$ lies between 0.01 and 0.04. Moreover, the average number of sub-intervals is smaller in case of hierarchical segmentation than sequential segmentation as shown in Fig. 6b. However, for reduction threshold values $\omega$ greater than 0.04, the sequential segmentation turns out to be the best approach for all investigated functions.

Table 2. Result of right-tailed and left-tailed two-sample $t$-test for pair-wise proposed algorithms (groups $G_1$, $G_2$ and $G_3$ correspond to binary, hierarchical and sequential, respectively.)

| $f(x)$ | $[x_0, x_0 + a)$ | Pair-wise test | Right-tailed $t$-test | Left-tailed $t$-test |
|---|---|---|---|---|
| $log(x)$ | $[0.625, 15.625)$ | $(G_1, G_2)$ | 0 | 0 |
|  |  | $(G_1, G_3)$ | 0 | 0 |
|  |  | $(G_2, G_3)$ | 0 | 0 |
| $e^x$ | $[0, 5)$ | $(G_1, G_2)$ | 0 | 0 |
|  |  | $(G_1, G_3)$ | 0 | 0 |
|  |  | $(G_2, G_3)$ | 0 | 0 |
| $tan(x)$ | $[-1.5, 0)$ | $(G_1, G_2)$ | 0 | 0 |
|  |  | $(G_1, G_3)$ | 0 | 1 |
|  |  | $(G_2, G_3)$ | 0 | 0 |
| $tanh(x)$ | $[-8, 0)$ | $(G_1, G_2)$ | 0 | 0 |
|  |  | $(G_1, G_3)$ | 0 | 1 |
|  |  | $(G_2, G_3)$ | 0 | 1 |
| $\dfrac{1}{1+e^{-x}}$ | $[-10, 0)$ | $(G_1, G_2)$ | 0 | 0 |
|  |  | $(G_1, G_3)$ | 0 | 1 |
|  |  | $(G_2, G_3)$ | 0 | 1 |
| $e^{\dfrac{-x^2}{2}}$ | $[-6, 0)$ | $(G_1, G_2)$ | 0 | 0 |
|  |  | $(G_1, G_3)$ | 0 | 1 |
|  |  | $(G_2, G_3)$ | 0 | 1 |

## 6 HARDWARE IMPLEMENTATION

Our second major contribution is the introduction of a generic, automatically synthesizable hardware implementation of the proposed approaches for the table-based function approximation. Fig. 7 depicts this hardware architecture.

The architecture's input is a bit vector $\Xi$ to be evaluated by the approximation of $f(x)$. The architecture's output is a bit vector $\Upsilon$. The input and the output are assumed to be user-defined as fixed point numbers represented by the tuples $(S^\Xi, W^\Xi, F^\Xi)$ and $(S^\Upsilon, W^\Upsilon, F^\Upsilon)$, respectively. Here, $S$ indicates the sign, $W$ corresponds to the length of the binary bit string, and $F$ denotes the number of bits used for the fractional part.

First, the input $\Xi$ passes through an interval selector unit determining the sub-interval containing $\Xi$ and consequently, the values of parameters specific to the sub-interval e.g., the spacing between breakpoints. Since the interval selector unit is implemented by using a comparator in each node of the binary tree generated from the set of sub-intervals, a single cycle implementation is not appropriate. Moreover, the sequential segmentation approach even generates an unbalanced binary tree, resulting in a generally larger set of cascaded comparators than the other two segmentation approaches. In our design flow, a pre-processing balancing step is therefore applied for any set $P$ of intervals that always delivers a balanced binary tree of comparators. Then, the address generator determines the addresses of the stored function values $A_i$ and $A_{i+1}$ corresponding to the breakpoints enclosing the input. These values are read from the BRAM. In the last stage, the subsequent block performs linear interpolation to determine $\Upsilon$.

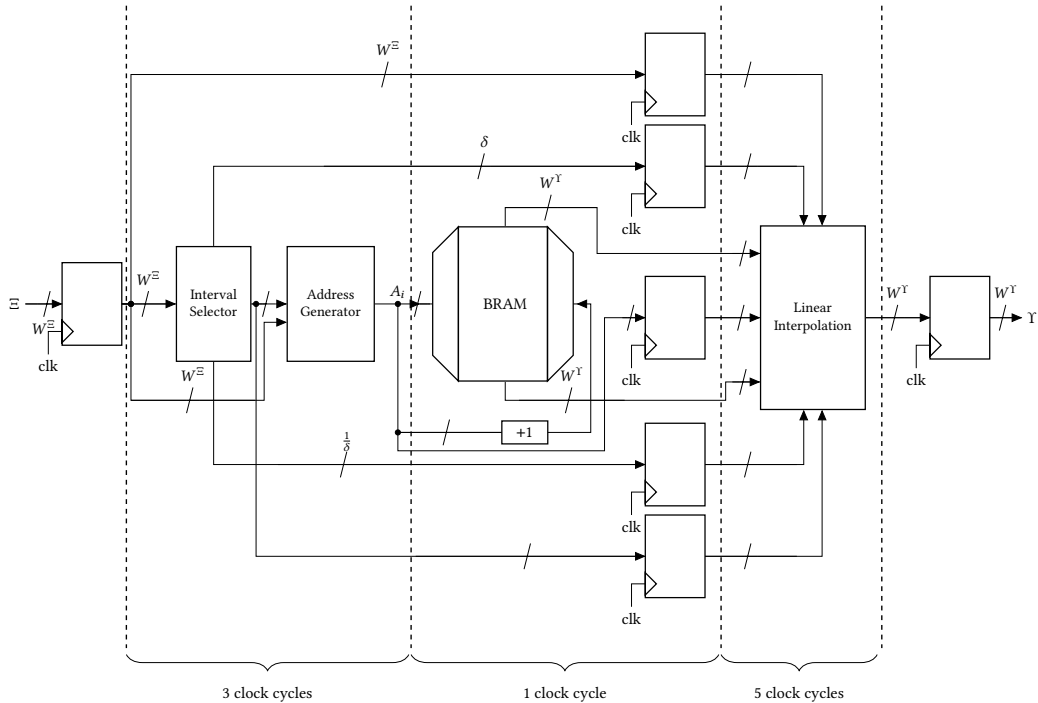

Fig. 7. Proposed generic hardware implementation for table-based function approximation using interval-splitting and BRAM instantiation. An input $\Xi$ of $W^\Xi$ bits in pre-specified fixed-point number format is evaluated. In just three clock cycles, the interval selector determines in which partition $\Xi$ is and its respective base address $A_i$ in the BRAM, as well as, a valid address is generated. In the next clock cycle, the two breakpoints required to evaluate $\Xi$ are looked up in the BRAM. Then, in another five clock cycles, the linear interpolation block calculates the approximated value of $f(x)$. The shown implementation is pipelined and has a latency of $L = 9$ clock cycles.

We implemented a design flow that automates the generation of the shown hardware architecture irrespective of the given function and interval. First, an approximation algorithm is selected and applied (see Sec. 5). The output $P$ of the algorithm is then used to generate a hardware description in VHDL. For the determination of the range values $y_i$ to be stored in BRAMs, we employ the HDL coder of Matlab [13] and adapt the code generation to instantiate BRAMs. The set $P$ is also directly used to implement the interval selector and the linear interpolation blocks. The arithmetic operations performed to compute the output are pipelined to increase the throughput of the circuit. The interval selector and address generator take three clock cycles together to generate valid address signals. The $y$ values are obtained from the BRAMs in the next clock cycle. Then, the pipelined linear interpolation block requiring five clock cycles produces the final output. Therefore, the latency ($L$) of the proposed architecture is constant at $L = 9$ per function evaluation. Note that this latency is independent of the function to be approximated, number formats, and number of sub-intervals determined by interval splitting. In the following, we evaluate the implementation of our segmentation approach in a target FPGA device to measure the memory footprint reductions, logic utilization (LUTs), BRAM utilization, and the operating frequency.

## 7 RESULTS

To evaluate the memory savings of the interval splitting approaches introduced in Sec. 5 also in real hardware, we again compare the even spacing table-based function approximation approach (*Reference*) against our newly introduced hierarchical interval segmentation table-based approximation approach. For the latter, we exemplarily consider the hierarchical approach. The goal of both the *Reference* approach and *Hierarchical segmentation* is to generate an efficient memory footprint function approximation of $f(x)$ for a given interval $[x_0, x_0 + a)$ and maximum absolute error $E_a$. Apart from comparing the memory footprint reductions $\Delta M_F$ obtained by each approach, we synthesized hardware implementations of both approaches and compared them regarding also the number of utilized BRAMs, the number of utilized LUTs, and the achievable operating frequency in MHz. Here, we used the Matlab/Simulink LUT Optimizer [14] to obtain the VHDL implementations of the *Reference* approach. For the purpose of comparison, we only customized the code delivered by Matlab's HDL coder to force the instantiation of the tabular function representation in BRAMs instead of using LUTs.

Regarding the *Hierarchical segmentation* approach, we utilized our newly introduced design flow (see Sec. 6) that automatically performs VHDL code generation and BRAM instantiation of the table-based function approximation. The benchmark considered for comparison consists of six test functions as presented in Table 3. We selected these functions because they present different gradient regions to examine the benefits of our proposed segmentation approach.

In the following, we will show that our proposed hierarchical segmentation-based approach is able to achieve fundamentally higher memory footprint reductions and efficient utilization of BRAMs over the *Reference* approach.

### 7.1 Test Setup

Our benchmark is composed of six test functions as presented in Table 3. Each function is evaluated by the *Reference* and the *Hierarchical segmentation* approach with a chosen absolute approximation error $E_a = 9.5367E - 07$ over the interval $[x_0, x_0 + a)$ presented in the second column in Table 3. The third and fourth columns in Table 3 show the input $(S^\Xi, W^\Xi, F^\Xi)$ and output $(S^\Upsilon, W^\Upsilon, F^\Upsilon)$ fixed-point format used to approximate the proposed test functions. Here, $S$ corresponds to the bit used to represent the sign, being 1 for a negative number and 0 for a positive. $W$ is the bit-width, and $F$ is the number of bits used for the fractional part.

As a target FPGA, we selected a Zynq-7000 Programmable System-on-Chip (PSoC) with 53, 200 LUTs, 106, 400 Flip-Flops, and up to 4.9 MB of BRAMs. We performed the synthesis of the circuits for the *Reference* and *Hierarchical segmentation* approaches using Vivado 2021.2. We synthesized six different circuits per approximated function for the hierarchical approach by varying the number of generated intervals $1 \leq n < 30$, where $n = |P| - 1$ stands for the number of generated intervals. For the trivial case of $n = 1$ (no splitting is performed), the results are equal to those delivered by the *Reference* approach.

### 7.2 Analysis of Synthesis Results

Fig. 8 presents the synthesis results obtained by the *Reference* and our approach *Hierarchical segmentation* regarding memory footprint ($M_F$), number of instantiated BRAMS, number of utilized LUTs, and operating frequency in MHz.

As can be seen, our proposed hierarchical approach is able to drastically reduce the memory footprint resulting in very efficient utilization of BRAMs (see Fig. 8a). Regarding logic utilization of the target FPGA, our hierarchical approach utilizes insignificantly more LUTs than the reference approach. This increase is due to the number of generated intervals impacting the size and depth of the binary tree of comparators of the interval selector (see Sec. 5). However, this typically represents only a 3 % overall utilization of the LUTs available in the target FPGA. Finally, our proposed approach delivers circuits ranging between 86.5 MHz and 88.5 MHz in terms of operating frequency (see Fig. 8b).

*7.2.1 Memory Footprint and BRAMs Utilization.* Fig. 8a presents the memory footprint, and the number of utilized BRAMs colored in blue and green, respectively, for the *Reference* (n=1) and the *Hierarchical segmentation* approaches. Here, we generated six implementations using the hierarchical approach for a varying number of generated intervals $n$ for each explored function. For example, when approximating the function $f(x) = tan(x)$, we generated six implementations with $n \in \{1, 3, 5, 13, 17, 29\}$ intervals each. The calculation of the memory footprint obtained by the *Reference* approach ($n = 1$) is according to Eq. (12) and Eq. (13) for *Hierarchical segmentation*.

At first, we can observe a considerable decrease in the memory footprint and the number of used BRAMs when we apply the hierarchical segmentation approach ($n > 1$) for all six considered test functions. The sixth and the seventh columns in Table 3 present the memory footprint ($\Delta M_F$) and the utilized BRAM ($\Delta BRAMs$) reductions obtained by our approach *Hierarchical segmentation* compared to *Reference* for each obtained partition. The memory footprint reduction was calculated according to Eq. (14).

In general, we can observe that as the number of obtained sub-intervals increases, the memory footprint decreases as well as the number of utilized BRAMs. E.g., for $f(x) = tan(x)$, the *Reference* results in a table with a $M_F^R = 81,543$ entries stored in 95 allocated BRAMs. On the other hand, the table obtained by our approach for a partition with $n = 3$ intervals resulted in a memory footprint reduction $\Delta M_F = 75\%$ and BRAM reduction $\Delta BRAMs = 66\%$. For $n = 5$, the corresponding memory footprint and BRAMs usage reduction resulted in $\Delta M_F = 80\%$ and $\Delta BRAMs = 83\%$, respectively. However, for $n = 13$, the memory footprint reduced by $\Delta M_F = 89\%$ but the $\Delta BRAMs = 83\%$ remained the same as for $n = 5$. Intuitively, we would expect a reduction in the BRAM usage as the memory footprint reduces. Nevertheless, this might not always be the case because of the storage capacity of each BRAM and how the data is internally stored.

According to the Xilinx 7-series specification, it is important to note that each BRAM can store up to 1,024 entries for an output bit-width of $W = 32$ bits. The number of address bits of one BRAM is thus 10. Similarly, for any memory footprint $M_F$, the number of required address bits is $\lceil log_2(M_F) \rceil$. Hence, the number of BRAMs of depth 1,024 ($2^{10}$) required to store $M_F$ values is given by $\frac{2^{\lceil log_2 M_F \rceil}}{1024} = 2^{\lceil log_2 M_F \rceil - 10}$. For example, let's consider two circuits synthesized to approximate $f(x) = tan(x)$ for $n = 5$ and $n = 13$ intervals. The reported memory footprints are $M_F = 15,644$ and $M_F = 8,798$, respectively. The number of address bits for both the cases is 14 and therefore, the number of allocated BRAMs is $2^{14-10} = 16$ for both implementations despite the large difference in memory footprints.

Also for the other test functions, we obtained significant memory footprint reductions and efficient BRAM utilizations. For example, for $f(x) = log(x)$, we reported memory footprint reductions ranging from 66 % to 85 % and BRAMs reduction ranging from 75 % to 87.5 %. In case of $f(x) = e^x$, the memory footprint reductions $\Delta M_F$ ranged from 38 % to 61 % with a BRAMs reduction $\Delta BRAMs$ of 50 %. For $f(x) = tan(x)$, the $\Delta M_F$ ranged from 57 % to 70 % with a BRAMs reduction $\Delta BRAMs$ up to 75 %. For $f(x) = e^{-\frac{x^2}{2}}$, the $\Delta M_F$ ranged from 40 % to 60 % with a BRAMs reduction $\Delta BRAMs$ up to 75 %. Finally, for $f(x) = \frac{1}{1+e^x}$, the $\Delta M_F$ ranged from 42 % to 55 % with a BRAMs reduction $\Delta BRAMs$ up to 75 %.

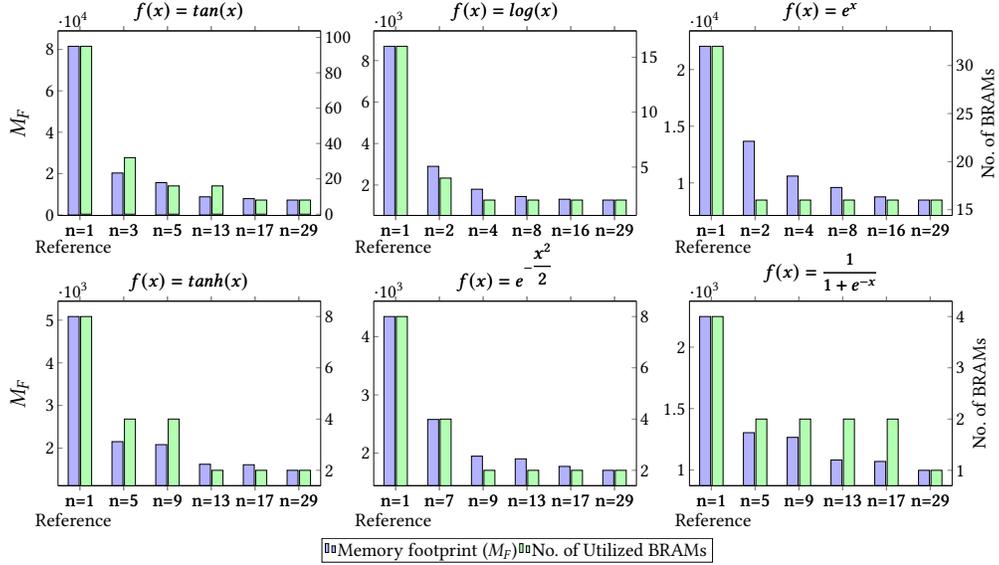

(a) Memory footprint and BRAM utilization

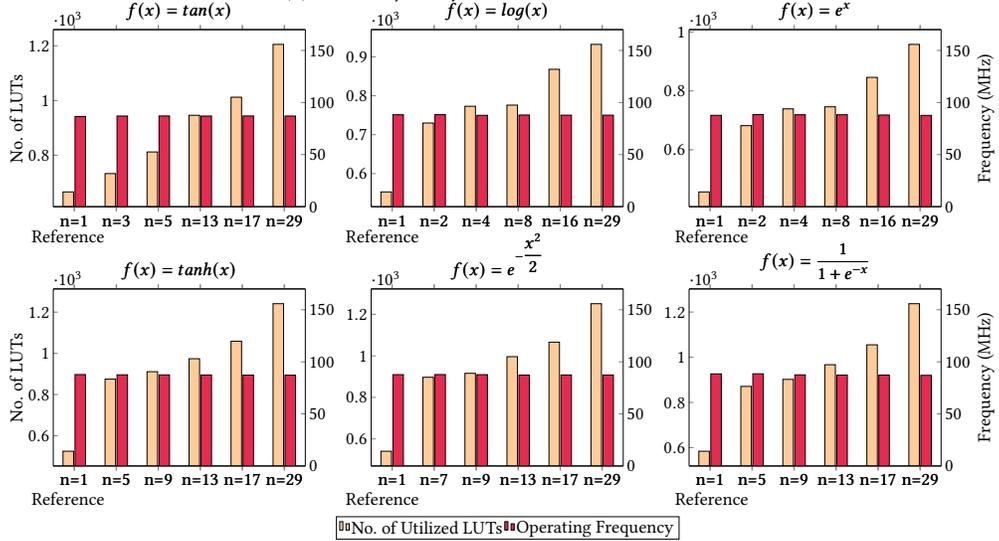

(b) LUT Utilization and Operating Frequency

Fig. 8. Synthesis results obtained using our *hierarchical segmentation* approach for the approximation of the six benchmark functions in Table 3 against the number of generated sub-intervals ($1 \leq n < 30$). In (a), we can observe that as the number of sub-intervals $n$ increases, so do the reductions in the memory footprint and the number of utilized BRAMs. In (b), the number of utilized LUTs and operating frequency are presented. Here, the number $n$ of intervals affects the number of utilized LUTs which are used to implement the interval selector. Finally, we can observe that the operating frequency is almost constant with $\sim 87$ MHz.

*7.2.2 LUTs Utilization.* In Fig. 8b, the light orange bars present the number of LUTs utilized for the six implementations obtained by the proposed *Hierarchical segmentation* approach ranging the number of generated intervals from $n = 1$ to 30. Here, the *Reference* corresponds to the case with no partition ($n = 1$). We can observe that the number of utilized LUTs increases with the number of generated intervals $n$ for all the presented benchmark functions. The increment in LUTs can be attributed to an increase in the number of comparisons required to traverse the binary tree in the interval selection block (see Fig. 7), which are tightly related to the number of generated intervals $n$ obtained by our proposed hierarchical approach. However, for all values $n$ of intervals, the overall overhead is typically less than 3 % of the available LUTs in the target FPGA.

Table 3. Evaluation benchmark composed of six functions and their characteristics presented from column two to four. Columns fifth to seventh present the memory footprint reduction $\Delta M_F$, the BRAM utilization reduction $\Delta BRAMs$ and the increment of LUTs $\Delta LUTs$ compared to the *Reference* approach in %.

| $f(x)$ | $[x_0, x_0 + a)$ | $(S^\Xi, W^\Xi, F^\Xi)$ | $(S^\Upsilon, W^\Upsilon, F^\Upsilon)$ | $n$ | $\Delta M_F$ [%] | $\Delta BRAMs$ [%] | $\Delta LUTs$ [%] |
|---|---|---|---|---|---|---|---|
| $tan(x)$ | $[-1.5, 1.5)$ | $(1, 32, 30)$ | $(1, 32, 27)$ | 3 | 75 % | 66 % | 10 % |
| | | | | 5 | 80 % | 83 % | 21 % |
| | | | | 13 | 89 % | 83 % | 42 % |
| | | | | 17 | 90 % | 91 % | 52 % |
| | | | | 29 | 91 % | 91 % | 81 % |
| $log(x)$ | $[0.625, 15.625)$ | $(0, 32, 28)$ | $(1, 32, 29)$ | 2 | 66 % | 75 % | 32 % |
| | | | | 4 | 79 % | 87.5 % | 39 % |
| | | | | 8 | 83 % | 87.5 % | 40 % |
| | | | | 16 | 84 % | 87.5 % | 57 % |
| | | | | 29 | 85 % | 87.5 % | 68 % |
| $e^x$ | $[0, 5)$ | $(0, 32, 29)$ | $(0, 32, 24)$ | 2 | 38 % | 50 % | 49 % |
| | | | | 4 | 51 % | 50 % | 61 % |
| | | | | 8 | 56 % | 50 % | 63 % |
| | | | | 16 | 60 % | 50 % | 85 % |
| | | | | 29 | 61 % | 50 % | 109 % |
| $tanh(x)$ | $[-8, 8)$ | $(1, 32, 27)$ | $(1, 32, 31)$ | 5 | 57 % | 50 % | 66 % |
| | | | | 9 | 59 % | 50 % | 73 % |
| | | | | 13 | 68 % | 75 % | 85 % |
| | | | | 17 | 68 % | 75 % | 102 % |
| | | | | 29 | 70 % | 75 % | 136 % |
| $e^{-\frac{x^2}{2}}$ | $[-6, 6)$ | $(1, 32, 28)$ | $(1, 32, 32)$ | 7 | 40 % | 50 % | 66 % |
| | | | | 9 | 55 % | 75 % | 69 % |
| | | | | 13 | 56 % | 75 % | 84 % |
| | | | | 17 | 59 % | 75 % | 97 % |
| | | | | 29 | 60 % | 75 % | 131 % |
| $\frac{1}{1 + e^x}$ | $[-10, 10)$ | $(1, 32, 27)$ | $(0, 32, 32)$ | 5 | 42 % | 50 % | 49 % |
| | | | | 9 | 43 % | 50 % | 54 % |
| | | | | 13 | 51 % | 50 % | 65 % |
| | | | | 17 | 52 % | 50 % | 81 % |
| | | | | 29 | 55 % | 75 % | 111 % |

7.2.3 *Operating Frequency.* In Fig. 8b, the red bars show the operating frequency reached by six implementations of the proposed *hierarchical segmentation* approach by varying the number of generated intervals $n$ between 1 to 30. The architecture is fully pipelined with a data introduction interval of only one clock cycle to start subsequent function evaluations. We can generally observe that the operating frequency lies between 86.5 MHz to 88.5 MHz for the six analyzed functions. For all test functions, the the critical path lies in the linear interpolation unit. A multiplication is carried out by two cascaded DSPs (Digital Signal Processor) which introduce an almost constant delay for all implementations. Slight differences (typically less than 2 MHz) are caused just by minor variations in the net and routing delays from design to design. Our proposed hierarchical approach reaches an overall average operating frequency of 87.5 MHz. Accordingly, the hardware implementation is able to produce an approximated function evaluation within $\frac{9\ clock\ cycles}{87.5\ MHz} \approx 102.8\ ns$.

## 8 CONCLUSION

In this article, we investigated an efficient way to approximate elementary functions given an interval and a maximum absolute error using interval splitting. First, we realized that by splitting the given interval into a set of sub-intervals and assuming a coarser sampling grid for low gradient regions, while respecting the given maximum approximation error $E_a$. Second, we proposed a generic hardware architecture to automatically synthesize such interval-split tabular function approximators with an evaluation latency of just $L = 9$ clock cycles per function evaluation. For the first time, we exploited explicitly the use of BRAMs by automatically inferring them in our design flow during the code generation of the hardware description. In consequence, reductions in the memory footprint and BRAM usage were shown to be achievable by our proposed approach. As future work, we want to explore more efficient packing of BRAMs and alternative sub-interval determination algorithms.

## REFERENCES

[1] Ben Adcock, Simone Brugiapaglia, and Clayton G. Webster. 2017. *Compressed Sensing Approaches for Polynomial Approximation of High-Dimensional Functions*. Springer International Publishing, 93–124.

[2] Mohammad Ahmadinejad, Mohammad Hossein Moaiyeri, and Farnaz Sabetzadeh. 2019. Energy and area efficient imprecise compressors for approximate multiplication at nanoscale. *International Journal of Electronics and Communications* 110 (2019), 152859.

[3] Ray Andraka. 1998. A Survey of CORDIC Algorithms for FPGA Based Computers. In *Proceedings of the 1998 ACM/SIGDA Sixth International Symposium on Field Programmable Gate Arrays (FPGA '98)*. Association for Computing Machinery, 191–200.

[4] A. Becher, J. Echavarria, D. Ziener, S. Wildermann, and J. Teich. 2016. A LUT-Based Approximate Adder. In *Proceedings of FCCM 2016*. 27–27.

[5] M. Brand, M. Witterauf, F. Hannig, and J. Teich. 2019. Anytime instructions for programmable accuracy floating-point arithmetic. In *Proceedings of ACM Int. Conf. on Comp. Fronts. (CF), 2019*. 215–219.

[6] Jon T. Butler, C.L. Frenzen, Njuguna Macaria, and Tsutomu Sasao. 2011. A fast segmentation algorithm for piecewise polynomial numeric function generators. *J. Comput. Appl. Math.* 235, 14 (2011), 4076–4082. https://doi.org/10.1016/j.cam.2011.02.033

[7] Linbin Chen, Jie Han, Weiqiang Liu, and Fabrizio Lombardi. 2017. Algorithm and Design of a Fully Parallel Approximate Coordinate Rotation Digital Computer (CORDIC). *IEEE Transactions on Multi-Scale Computing Systems* 3, 3 (2017), 139–151.

[8] Davide De Caro, Nicola Petra, and Antonio G. M. Strollo. 2011. Efficient Logarithmic Converters for Digital Signal Processing Applications. *IEEE Transactions on Circuits and Systems II: Express Briefs* 58, 10 (2011), 667–671. https://doi.org/10.1109/TCSII.2011.2164159

[9] Hongxi Dong, Manzhen Wang, Yuanyong Luo, Muhan Zheng, Mengyu An, Yajun Ha, and Hongbing Pan. 2020. PLAC: Piecewise Linear Approximation Computation for All Nonlinear Unary Functions. *IEEE Transactions on Very Large Scale Integration (VLSI) Systems* 28, 9 (2020), 2014–2027. https://doi.org/10.1109/TVLSI.2020.3004602

[10] Jorge Echavarria, Stefan Wildermann, Andreas Becher, Jürgen Teich, and Daniel Ziener. 2016. FAU: Fast and Error-optimized Approximate Adder Units on LUT-Based FPGAs. In *Proceedings of Field-Programmable Technology (FPT)*. 213–216. https://doi.org/10.1109/FPT.2016.7929536

[11] J. Han and M. Orshansky. 2013. Approximate computing: An emerging paradigm for energy-efficient design. In *2013 18th IEEE European Test Symposium (ETS)*. 1–6.

[12] Dong-U Lee, Ray C. C. Cheung, Wayne Luk, and John D. Villasenor. 2009. Hierarchical Segmentation for Hardware Function Evaluation. *IEEE Transactions on Very Large Scale Integration (VLSI) Systems* 17, 1 (2009), 103–116. https://doi.org/10.1109/TVLSI.2008.2003165

[13] MATLAB. 2019. *version 9.6.0 (R2019a)*. The MathWorks Inc., Natick, Massachusetts.

[14] MATLAB. 2021. *Optimize Lookup Tables for Memory-Efficiency Programmatically*. The MathWorks Inc. https://de.mathworks.com/help/fixedpoint/ug/optimize-lookup-tables-for-memory-efficiency-programmatically.html

[15] Jean-Michel Muller. 1999. A Few Results on Table-Based Methods. *Reliable Computing* 5 (1999), 279–288.

[16] Srinivasan Narayanamoorthy, Hadi Asghari Moghaddam, Zhenhong Liu, Taejoon Park, and Nam Sung Kim. 2015. Energy-Efficient Approximate Multiplication for Digital Signal Processing and Classification Applications. *IEEE Transactions on Very Large Scale Integration Systems* 23, 6 (2015), 1180–1184. https://doi.org/10.1109/TVLSI.2014.2333366

[17] Hyoju Seo, Yoon Seok Yang, and Yongtae Kim. 2020. Design and Analysis of an Approximate Adder with Hybrid Error Reduction. *Electronics* 9, 3 (2020).

[18] David I Shuman, Pierre Vandergheynst, and Pascal Frossard. 2011. Chebyshev polynomial approximation for distributed signal processing. In *2011 International Conference on Distributed Computing in Sensor Systems and Workshops (DCOSS)*. 1–8.

[19] Ye Tian, Ting Wang, Qian Zhang, and Qiang Xu. 2017. ApproxLUT: A novel approximate lookup table-based accelerator. In *2017 IEEE/ACM International Conference on Computer-Aided Design (ICCAD)*. 438–443.

[20] Xilinx. 2019. *7 Series FPGA Memory Resources*. https://www.xilinx.com/. https://www.xilinx.com/support/documentation/user_guides/ug473_7Series_Memory_Resources.pdf

[21] Q. Xu, T. Mytkowicz, and N. Kim. 2015. Approximate computing: A survey. *IEEE Design & Test* 33, 1 (2015), 8–22.